\documentclass[10pt, conference]{IEEEtran}

\usepackage{graphicx}
\usepackage{multirow}
\usepackage{threeparttable}
\usepackage{listings}
\usepackage{framed}
\usepackage{url}
\usepackage{color}
\usepackage{tabularx}
\usepackage{booktabs}
\usepackage{makecell}
\usepackage{textcomp}
\usepackage{array}
\usepackage{amsmath}
\usepackage[linesnumbered, vlined, ruled]{algorithm2e}
\usepackage{mathrsfs}
\usepackage{enumitem}
\usepackage{verbatimbox}
\usepackage{setspace}
\usepackage{cite}

\def\BibTeX{{\rm B\kern-.05em{\sc i\kern-.025em b}\kern-.08em
    T\kern-.1667em\lower.7ex\hbox{E}\kern-.125emX}}

\definecolor{dkgreen}{rgb}{0,0.6,0}
\definecolor{gray}{rgb}{0.5,0.5,0.5}
\definecolor{mauve}{rgb}{0.58,0,0.82}

\begin{document}

    \title{Improving Tese Case Generation for Python Native Libraries Through Constraints on Input Data Structures}
    
    \author{
        Xin Zhang$^{1, 3}$, 
        Xutong Ma$^{1, 3}$, 
        Jiwei Yan$^{2, 3}$, 
        Baoquan Cui$^{1, 3}$, 
        Jun Yan$^{1, 2, 3}$, 
        Jian Zhang$^{1, 3, \dag}$\thanks{$^{\dag}$Corresponding author} \\
    
        $^1$ State Key Laboratory of Computer Science, Institute of Software, Chinese Academy of Sciences \\
    
        $^2$ Technology Center of Software Engineering, Institute of Software, Chinese Academy of Sciences \\
    
        $^3$ University of Chinese Academy of Sciences \\
    
        Email: \{zhangxin19, maxt, yanjw, cuibq, yanjun, zj\}@ios.ac.cn
    }

    \maketitle
    
    \pagestyle{plain}
    \thispagestyle{plain}

    \begin{abstract}
        Modern Python projects execute computational functions using native libraries and give Python interfaces to boost execution speed; hence, testing these libraries becomes critical to the project's robustness.
One challenge is that existing approaches use coverage to guide generation, 
but native libraries run as black boxes to Python code with no execution information.
Another is that dynamic binary instrumentation reduces testing performance as it needs to monitor both native libraries and the Python virtual machine.

To address these challenges, in this paper,
we propose an automated test case generation approach that works at the Python code layer.
Our insight is that many path conditions in native libraries are for processing input data structures through interacting with the VM.
In our approach, we instrument the Python Interpreter to monitor the interactions between native libraries and VM, 
derive constraints on the structures, 
and then use the constraints to guide test case generation.
We implement our approach in a tool named PyCing and apply it to six widely-used Python projects.
The experimental results reveal that with the structure constraint guidance, 
PyCing can cover more execution paths than existing test cases and state-of-the-art tools.
Also, with the checkers in the testing framework \textit{Pytest}, 
PyCing can identify segmentation faults in 10 Python interfaces and memory leaks in 9.
Our instrumentation strategy also has an acceptable influence on testing efficiency.
    \end{abstract}


    \section{Introduction}
        \label{sec: introduction}
        Benefitting from its flexible syntax and vast third-party libraries, 
Python has become the dominating programming language in many popular fields, 
including machine learning, scientific computation, and image processing~\cite{javed2019analysis}. 
Projects develop calculating functions utilizing native libraries and give Python code with interfaces to fulfill the performance expectations in these fields.
These native libraries are mainly written with statically typed languages such as C, C++, FORTRAN, and Cython~\cite{cython} and made available to Python code through dynamic link libraries.
Many high-performance computing projects have developed their own native libraries, 
like \textit{NumPy}~\cite{NumPy} and \textit{SciPy}~\cite{SciPy} for scientific computing, 
and \textit{Sklearn}~\cite{Sklearn} for machine learning.
In this paper, we refer to a Python interface provided by native libraries as a \textbf{native method},
and refer to the type and attributes of one Python object as the \textbf{structures} of the Python object.

Because data structures change across programming languages, defect detection mechanisms in Python, 
such as automatic memory management, cannot cover these libraries.
While utilizing native libraries has several advantages, 
such as better execution performance and interface with the operating system, 
they may also result in fatal errors,
such as type conflict~\cite{HuZ20, Khan21, chen2020empirical}, 
performance loss~\cite{ZhangS21, zhou2020harp}, 
and dependency issues~\cite{MukherjeeAR21, ShafferCT21}.
Besides, native libraries run on the operating system, not the Python virtual machine.
Some defects in native libraries may cause VM crashes, like segmentation faults. 
The rising popularity of native libraries emphasizes the critical nature of their testing.

Fuzzing methods are effective in detecting bugs under the guidance of code coverage~\cite{LukasczykKF20, CadarDE08, DerezinskaH14}.
They get code coverage by instrumenting the source code, 
which is not difficult in cases when just focusing on one programming language.
However, they fail in instrumenting the native libraries in real-world Python projects since they cannot parse dynamic link libraries and are incompatible with the complicated compilation process.
One option is dynamic binary instrumentation.
Because the Python virtual machine is essential for running native libraries, 
this option needs to instrument both the VM and native libraries, 
which can significantly lower test efficiency.
Moreover, invoking native libraries without a VM necessitates rewriting all VM-related functions, 
which is time-consuming and labor-intensive~\cite{WeiLOCZ18}. 

Some approaches for other languages, such as Java, test native libraries through summaries~\cite{Lee19, ZhaoWLPKW19, HwangLKR21, KimHK19}.
They employ statically analyzing C code to capture constraints in summaries, which they subsequently use to guide test case generation. Because Java and native libraries are both statically typed languages, 
the input data structures are known, so these approaches only need to focus on the input value.
However, building summaries for Python native libraries can be challenging.
On the one hand, native libraries in Python have multiple languages, including C, C++, FORTRAN, and Cython.
Combining summaries for different languages is complex.
On the other hand, Python does not limit or validate input types to native methods due to the dynamically typed language feature.
To provide the same functionality for input data of varying types and attributes, 
native libraries include lots of path conditions related to structures~\cite{Peng0H21}. 
Building summary necessitates both modeling functions for processing input data structures 
and dealing with the situation where native libraries call Python code.
It poses considerable difficulties in producing effective summaries for testing Python native libraries.

Compared with other test scenarios, testing Python native libraries has two key challenges.
\begin{itemize}[topsep=1pt, partopsep=1pt, leftmargin=15pt]

    \item 
    \textbf{Lacking information about structure constraints in native libraries.}
    If there is no range of structures that native libraries can handle,
    it can be hard to improve test effectiveness by randomly generating input data with Python built-in types. 
    Besides, going through all the common types and attributes will result in a combinatorial explosion.

    \item 
    \textbf{Lacking execution information of native libraries.}
    Execution information is critical not just for test case generation but also for evaluation.
    Native libraries operate as black boxes for Python code, offering no execution information.
    Moreover, dynamic binary instrumentation will dramatically degrade test efficiency.
 
\end{itemize}

In this paper, we propose an automated test case generation approach 
guided by structure constraints in Python native libraries.
Given the overhead of directly invoking native libraries, 
we choose to test native libraries from the Python code layer.
Our insight is that many path conditions in native libraries are for input data structures, 
and they must process the structures using API functions provided by the Python Interpreter.
Thus, we can track the calls to these APIs when executing the native libraries, 
and build structure constraints based on the API call logs.
We chose to instrument these API functions to address the first challenge.
In this way, we can retrieve the arguments and return values every time the native libraries use these functions, 
then utilize the outputs to build constraints on input data structures.
To address the second challenge, 
as the structure constraints are part of path conditions in native libraries,
they can reflect the number of paths explored by test cases.
Our approach takes the source code of a project as input.
We first retrieve native methods from the documentation.
Then, we execute these methods to build structure constraints for existing input data.
Following that, constraints are reversed and used to generate new input data. 
We can gradually explore paths in native libraries by utilizing newly created objects as input data.

We implement our approach in a tool named PyCing(\textbf{Py}thon/\textbf{C} Test\textbf{ing}).
To demonstrate the benefits of our approach, we apply it to native libraries in 6 popular open-source Python projects.
We first compare PyCing with test cases in the projects.
Because existing tools cannot obtain coverage from native libraries,
we use a set of handwritten codes as benchmarks to successfully run these tools and compare them with PyCing.
The experimental results show that PyCing can cover more branch conditions relating to structures than test cases.
Moreover, objects from PyCing can identify 10 segmentation errors, 26 known memory leaks, and 9 novel leaks using checkers from the popular testing framework \textit{Pytest}~\cite{pytest}.
In the comparisons to two state-of-the-art Python testing tools, 
the findings indicate that PyCing can generate objects with different structures, 
and these objects can be treated as seeds for other tools.

In conclusion, we make the following contributions.
\begin{itemize}[topsep=1pt, partopsep=1pt, leftmargin=15pt]
    \item 
    We propose an instrumentation strategy 
    to collect constraints on input data structure from native libraries.
    Compared with existing instrumentation strategies, 
    ours can be applied to large Python projects and has less impact on the test efficiency.
    
    \item
    We propose an automated test case generation approach based on structure constraints, 
    and we implement it in a tool named PyCing.
    To the best of our knowledge, it is the first tool for automatically testing Python native libraries.
    
    \item  
    PyCing is applied to 6 real-world Python projects, 
    and the results indicate that it can explore more paths than test cases in the projects.
    Input data from PyCing can detect segmentation faults in 10 native methods and memory leaks in 9.

\end{itemize}

    \section{Background and Motivation}
        First, we will briefly introduce the Python execution environment and how Python code interacts with native libraries.
Following that, an example from the \textit{NumPy} project will be used to demonstrate the significance of input data structures in testing native libraries.

\subsection{The Python Execution Environment}
    \label{sec: environment}
    Python is an interpreted language with many interpreters available, 
    including \textit{CPython}~\cite{CPython} written in C/C++, 
    \textit{Intel-Python}~\cite{IntelPython} developed by Intel, 
    and \textit{JPython}~\cite{Jython} written in Java.
    Because \textit{CPython} is the official version and is commonly used in the development, 
    all the keywords \textit{Python Interpreter} and \textit{interpreter} in this paper refer to \textit{CPython}.

    The interpreter executes Python code on a stack-based virtual machine 
    and saves the stack frame throughout execution using Python objects called \texttt{PyFrame}.
    Each \texttt{PyFrame} provides access to the execution context, 
    including the current local and global variables, the line of executed code, and the location of the Python file.
    Users can acquire \texttt{PyFrame}s by registering a tracer to the code.
    Many Python debug tools, like \textit{pdb}~\cite{pdb}, employ this tracer mechanism.

\subsection{Python Native Libraries}
    \label{sec: interaction}
    Developers can compile functions developed by other programming languages into dynamic link libraries, 
    which can subsequently be invoked from Python code.
    To establish correlations between Python methods and native functions for analysis,
    some researchers analyze the source code of native libraries statically~\cite{FourtounisTS20, ZhaoWLPKW19, HuZHX21, abs-2204-08237}.
    Because developers can specify the names of Python interfaces in many ways, 
    these approaches are inaccurate when applied to real-world projects, 
    making call graphs between Python code and native libraries hard to construct.

    A Python object is handled in the native libraries as a C struct \texttt{PyObject}, 
    where types and attributes of the object correspond to the fields in the \texttt{PyObject}.
    Thus, in this paper, we regard all the operations on a \texttt{PyObject} in native libraries 
    as operations on the attributes and types of the corresponding Python object.

    To help native libraries manage Python objects and access the Python virtual machine,
    the interpreter provides a set of C/C++ functions named Python/C API~\cite{pythoncapi}.
    These API functions enable native libraries to check types, extract attributes, 
    allocate and free memory in the Python heap area, and throw exceptions to Python code.
    
\subsection{A Motivating Example}
    \label{sec: memory leak} 
    The interpreter manages the lifetime of objects with a reference counting mechanism, 
    relieving developers of the work of manually freeing objects and reducing the danger of memory leaks~\cite{memmanage}.
    But this mechanism demands developers to manage memory manually while processing Python objects in native libraries.
    Thus, if the reference count of a Python object is increased in native libraries, 
    the interpreter will not free it even though it is no longer in use~\cite{LiT14, MaoCXS16}.
    
    Figure~\ref{motivating} shows one of the C functions in the implementation of a native method \texttt{numpy.power}.
    This native method takes two Python objects as input and calculate their powers.
    The presented C function receives one \texttt{PyObject} 
    and one double value as input and returns the power value through the second argument, 
    where argument \texttt{o2} points to the second argument of this native method.
    In line 4, this function uses a Python/C API function \texttt{PyLong\_Check} to check if \texttt{o2} is of type \texttt{int}.
    On line 9, the API function \texttt{PyFloat\_Check} checks for the type \texttt{float}.
    API function \texttt{PyIndex\_Check} at line 13 checks if \texttt{o2} includes a member method \texttt{\_\_index\_\_}.
    Then at line 14, this function invoke the member method \texttt{\_\_index\_\_} through API function \texttt{PyNumber\_Index}.
    \begin{figure}[h]
  \centering
  \begin{lstlisting}
static NPY_SCALARKIND is_scalar_with_conversion(
    PyObject *o2, double* out_exponent){
  ...
  if(PyLong_Check(o2)){
    ...
    return NPY_NOSCALAR;
  }
  ...
  if(PyFloat_Check(o2)){
    ...
    return NPY_FLOAT_SCALAR;
  }  
  else if(PyIndex_Check(o2)){
    PyObject* value = PyNumber_Index(o2);
    ...
    return NPY_INTPOS_SCALAR;
  }
  ...
 }
  \end{lstlisting}
  \caption{One of C functions in the implementation of the native method \texttt{numpy.power}.}  
  \label{motivating}
\end{figure}

    The API function \texttt{PyNumber\_Index} will raise the return value's reference count to prevent this value from being recycled throughout the caller's procedure.
    A leak will occur if the native method finishes without decreasing the reference count.
    To trigger this leak, the second argument of the native method must 
    make the API functions on lines 4 and 9 return \texttt{False} and the API function on line 13 return \texttt{True}. 
    Thus, the input data cannot inherit from the types \texttt{int} and \texttt{float},
    and must have the method \texttt{\_\_index\_\_}. 
    
    An example of input data to trigger this leak is as follows, 
    in which we construct a class and inherit it from type \texttt{str}, 
    then add the needed method to this class from lines 2 to 4.
    In line 5, we instantiate this class and pass it a random string. 
    However, such input data does not exist in the project's test cases.
    Existing Python testing tools can only mutate values, 
    and they cannot generate test case to trigger this leak when their seeds do not contain such structures. 
    \begin{lstlisting}[language=Python]
class self_class(str):
  index = 1
  def __index__(self):  
    return self.index
obj2 = self_class('')
    \end{lstlisting}

    \section{Terminology}
        In this section, we will introduce some conceptions used in our approach.

\textbf{A Python object.}
A Python object can be described as a pair: 
$$ obj = \langle Type\,, Attrs \rangle $$, 
where $Type$ denotes the object's type and $Attrs$ is a collection of its attributes. 
One attribute can be a member method, a member variable, or an element in a collection type object.
Each attribute is a Python object as well.

\textbf{Constraints on type inheritance.}
Let set $T$ represent all Python types, 
and $ \forall t_1, t_2 \in T, t_1 \preceq t_2 $ denotes that types $t_1$ and $t_2$ are the same, 
or that type $t_1$ inherits from type $t_2$.

If two sets $ T_{bt} \subseteq T $ and $ T_{nbt} \subseteq T $ satisfy  
$ \forall t1 \in T_{bt}, t2 \in T_{nbt}, \lnot(t1 \preceq t2) \wedge \lnot(t2 \preceq t1) $ 
,we can define path constraints on the type inheritance of an object (Type constraints for short) as:
$$ \forall t1 \in T_{bt}, t2 \in T_{nbt}, (obj.Type \preceq t1) \wedge \lnot(obj.Type \preceq t2) $$

Set $T_{bt}$ represents all the types that an object should be or should inherit from, 
whereas the set $T_{nbt}$ contains all the types that should not be or should not inherit from

\textbf{Constraints on attribute existence.}
Let set $A$ represent all Python objects, 
and $ \exists a \in A, a \in obj.attrs $ indicates that Python object $ obj $ contains the attribute $ a $.

If there are two sets $ A_{bt} \subseteq A $ and $ A_{nbt} \subseteq A $, 
and they fulfill $ A_{bt} \cap A_{nbt} = \emptyset $, 
we can express the constraints on attribute existence of an object (attribute constraints for short) as:
$$ \forall a1 \in A_{bt}, a2 \in A_{nbt}, (a1 \in obj.Attrs) \wedge (a2 \notin obj.Attrs) $$

Set $A_{bt}$ represents all of the attributes that an object needs to contain, 
whereas the set $A_{nbt}$ represents the attributes that should not possess.

\textbf{Structure Constraint.}
Every Python object can be mapped to a quad $(T_{bt}, T_{nbt}, A_{bt}, A_{nbt})$ 
based on the definitions of constraints on type inheritance and attribute existence.

Because distinct attributes in a Python object correspond to different quads, 
one structure constraint of a Python object ($SC$ for short) is a conjunctive normal form of a set of mappings:
$$ SC_{object} = map_1 \wedge map_2 \wedge \cdots \wedge map_i $$.
Each $map_i$ is a mapping connections between an object and a quad:  
$ map_i = obj_i \mapsto (T_{i\_bt}, T_{i\_nbt}, A_{i\_bt}, A_{i\_nbt}) $,
where $obj_i$ can be the Python object $object$ or one of its attributes.

    \section{The Proposed Approach}
\begin{figure}[h]
    \centering  \includegraphics[width=\linewidth]{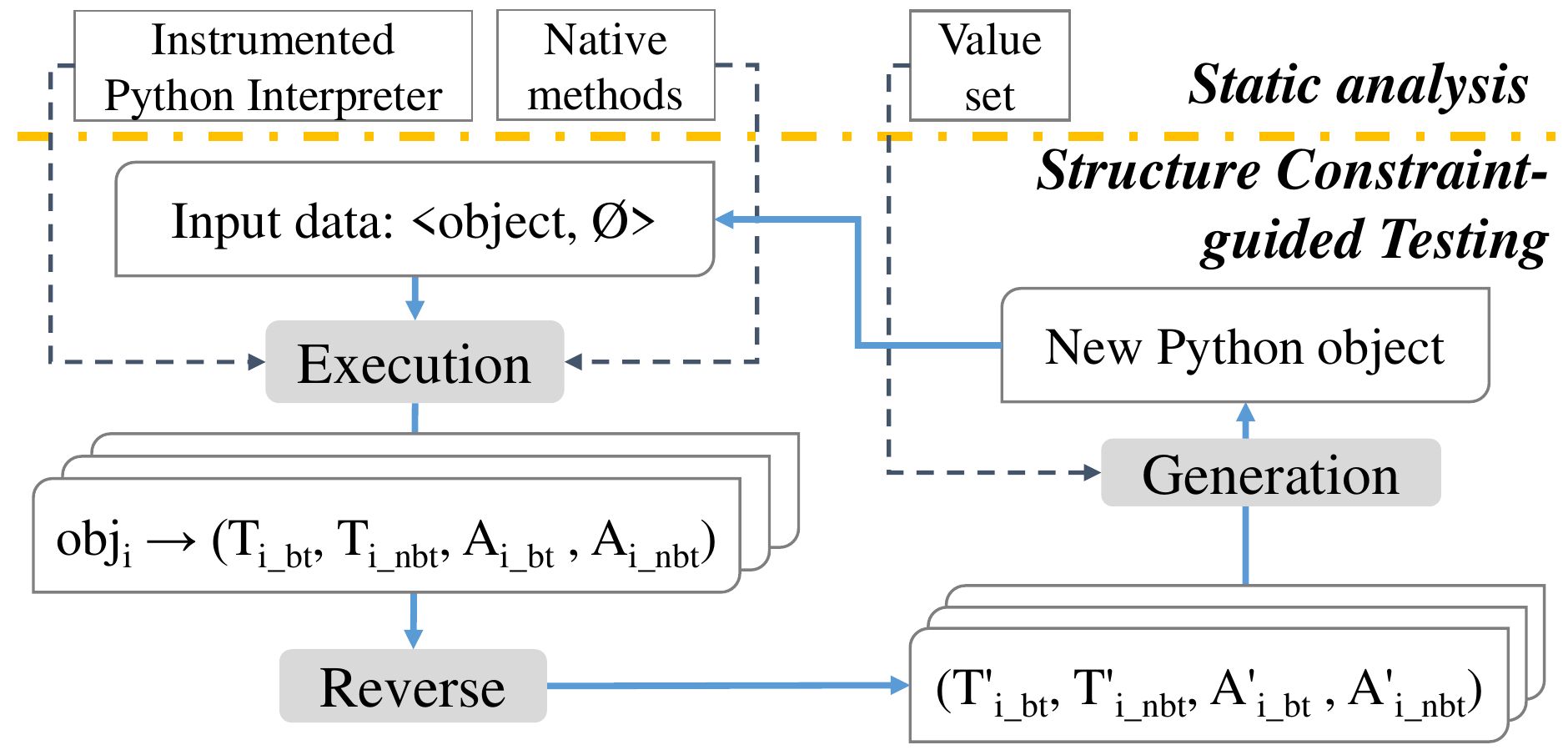}
    \caption{The workflow of our approach.}  \label{workflow}
\end{figure}

        Figure~\ref{workflow} depicts the workflow of our approach, 
encompassing two phases: 
\textbf{Static analysis} and \textbf{Structure constraint-guided Testing}.
In the first phase, we construct a value seed to house all of the values used in the projects.
Also, we instrument the Python/C API to capture execution information.
The second stage uses a loop to iteratively explore the execution paths, 
stopping when no new $SC$ appears.
In each loop, we construct $SC$s for existing input data based on API call logs.
Guided by the $SC$s, our approach generates new Python objects as input data for the next loop.

\subsection{Static analysis}
    This section illustrates how we collect project values and instrument the Python Interpreter.
    
    \subsubsection{Value Set Construction}
        \label{sec: valueset}
        In addition to $SC$s, there are constraints on input values, 
        such as branch conditions that compare the input value to a certain value.
        However, extracting value constraints from native libraries exactly can be problematic.
        On the one hand, extracting from source code requires complete call function graphs and path-sensitive interprocedural analysis, 
        which are challenging to achieve with present static analysis tools.
        On the other hand, it is tricky to relate the instructions to input data if monitoring the instructions.
        We cannot find vulnerabilities lying in paths if we generate values randomly.

        To circumvent value constraints, we apply a lightweight strategy, 
        where we collect all of the values used in the projects and employ them in the generating phase.
        We can produce values that satisfy such branch conditions in this manner.
        In detail, we collect values from two aspects:
        the literals in the source code and objects passed to native methods in the projects. 
        We first scan Python and native library source code and store all the literals.
        Then, to collect objects passed to the native methods, 
        we utilize the debug tool \textit{pdb} to insert breakpoints to each native method call.
        In this way, we can stop at each callsite and dump the input data while running the test suites in the projects. 
        We build the value set \textit{ValueSet} to hold all of the data we collect.

    \subsubsection{Python Interpreter Instrumentation}
        \label{sec: instrument}
        As mentioned in Section~\ref{sec: introduction}, 
        native libraries serve as black boxes for Python code.
        In our approach, we utilize Python/C APIs to monitor native library execution.
        We instrument these API functions at the source code level and recompile the interpreter.

        The interpreter supports roughly a thousand API functions, 
        with parts implemented using function-like macros. 
        Because not all the functions are for processing input data, 
        we manually review the documentation and filter two categories of functions: 
        one for type checking and another for attribute extracting.
        We utilize two separate instrumentation strategies to output in the same format 
        while their implementation is not the same.
        \begin{itemize}[topsep=1pt, partopsep=1pt, leftmargin=15pt]

            \item
            \textbf{API functions for type checking.} 
            This category of functions takes a reference as an input and returns a boolean value indicating whether the referred Python object is of the given type.
            We collect 50 functions from the documentation and insert code to output the reference, the verified type, and the result.
            
            \item
            \textbf{API functions for attribute extracting.}
            This category of API functions receives two pointers as input, 
            one to a Python object and the other to a key, 
            and returns the attribute if it exists or NULL if it does not. 
            We gather 14 functions and insert code to print the two pointers and the return value.

        \end{itemize}

        Using the output of the instrumented code, 
        we can abstract each API function call into a 4-tuple:  
        $ API(o, act, val_{act}, ret) $, 
        where $o$ stands for the input data, 
        $act$ for the functionality, such as type checking or attribute extracting, 
        $val_{act}$ for the argument, 
        and $ret$ for the return value.
        
        \textbf{Example for instrumentation:}
            From lines 2 to 6 in the following code, 
            we give the type-checking API function \texttt{PyDict\_Check} implementation, 
            where we instrument code at line 4 to output the checked type \texttt{dict}, 
            the checking result, and the argument.
            Besides, we present the simplified implementation of 
            API function \texttt{PyMapping\_GetItemString} from lines 8 to 13.
            This function is for extracting attributes from the input data.
            Lines 10-11 are the instrumented code that prints the attribute \texttt{r}, the key \texttt{key}, and the input data \texttt{o}.
            \begin{lstlisting}
// Objects/dictobject.c
int PyDict_Check(void *op){
  int result = PyType_FastSubclass(
                 PY_TYPE(op), Py_TPFLAGS_DICT_SUBCLASS);
  printf("%p, Chk, dict, %d \n", op, result);
  return result;
}
// Objects/abstract.c
PyObject * PyMapping_GetItemString(PyObject *o, char *key){
  PyObject * r = PyObject_GetItem(o, key);
  printf("%p, Get, %s, %p \n", o, key, r);
  return r;
}
\end{lstlisting}
            
\subsection{Execution}
    \label{sec: construction}
    In this part, we construct $SC$s for existing input data.
On a high level, we use an instrumented Python Interpreter to call native methods.
Then, using the interpreter output, we gather associated API function call logs and construct the $SC$s.
    
We will receive a series of API function calls after executing.
However, we cannot match the C structures processed by the API functions with the input data just based on logs.
This is because native libraries can also employ API functions to process locally created objects 
or one of the arguments passed to the native method.
Without knowing which calls are for input data, 
we will construct plenty of constraints that are not related to the input data, 
limiting our exploration's efficiency.
Consider the code shown in Figure~\ref{motivating},
API function calls to object \texttt{o2} are for this native method's second argument.
It will be hard to explore paths in this C function if we wrongly assume that these calls are for processing the first argument.

Thus, we scan the logs twice to build the $SC$s.
We gather type and attribute constraints on input data for the first time.
The second time, 
we collect constraints on attributes of input data to fulfill the $SC$s.

Because all Python objects are stored in Python heap space, 
an object's memory address will not change when transferred from Python code to native libraries.
Thus, we can filter the function calls that process the input data using the memory address.
Before invoking a native method, 
we first retrieve the memory addresses for the input data at the Python code layer, 
then append these addresses to the $SC$s.
If the address in one call appears in $SC$s, 
it indicates that this function is for processing the input data.
We can establish the type and attribute constraints for the input data in the first scan.

However, using memory address matching to create constraints for attributes is inefficient.
Because an attribute is also a Python object, 
the $Attrs$ of a Python object is tree-structured.
Acquisition of memory addresses for all the attributes can cause recursive traversal and consumes a lot of space.

As the addresses of input data were added to the $SC$s in the first step, 
we filter the calls of attribute extraction and add the address of the extracted attribute to the $SC$.
We can then use the extracted attribute to filter API function calls and progressively generate $SC$s for the input data.

\begin{table*}[h]
    \centering
    \renewcommand{\arraystretch}{1.5}
    \caption{Rules for constructing structure constraints.} 
    \begin{tabular}{m{0.5cm}m{3.3cm}m{3.5cm}m{1cm}m{7.2cm}}
        \toprule
        Rule  &  Example Statements  &  Call patterns  &  OP ID  &  Updating operations  \\
        
        \hline
        \multirow{2}{*}{R1}  &
        \multirow{2}{*}{PyDict\_Chk(obj);}  &
        \multirow{2}{*}{$API(obj, Chk, val_{act}, ret)$}  
              &  OP1.1  &  $ obj \mapsto (T_{obj\_bt} \cup \{val_{act}\}) | ret = True \wedge obj \in SC $  \\
        &  &  &  OP1.2  &  $ obj \mapsto (T_{obj\_nbt} \cup \{val_{act}\}) | ret = False \wedge obj \in SC $  \\

        \hline
        \multirow{3}{*}{R2}  &
        \multirow{3}{*}{PyList\_GetItem(obj, ind);}  &
        \multirow{3}{*}{$API(obj, Get, val_{act}, ret)$}   
              &  OP2.1  &  $ SC \wedge ( ret \mapsto () ) $  \\
        &  &  &  OP2.2  &  $ obj  \mapsto (A_{obj\_bt} \cup \{val_{act}\}) | ret \neq NULL \wedge obj \in SC $  \\
        &  &  &  OP2.3  &  $ obj  \mapsto (A_{obj\_nbt} \cup \{val_{act}\} ) | ret = NULL \wedge obj \in SC $  \\

        \bottomrule
    \end{tabular}
    \label{table: constructing}
\end{table*}
The two guidelines in Table~\ref{table: constructing} define the building process.
We present example statements, API function call patterns, and updating operations for each rule. 
When we process an API function for type checking, such as the call patterns in rule \textbf{R1}, 
we adjust the type constraints of $obj$ based on the return result if the $obj$ is recorded in the $SC$.
Specifically, if the return value is true, 
we add the $val_{act}$ to the set $T_{bt}$ through operation \textit{op1.1}, 
or to the set $T_{nbt}$ by operation \textit{op1.2}.
When dealing with an API function for attribute extracting, as described in rule \textbf{R2}, 
we modify the attribute constraints if the $obj$ is in the $SC$.
Firstly, we add the attribute with an empty set to the $SC$s using the operation \textit{op2.1}.
Then, similar to the operations in rule \textit{R1}, if the return result is not null, 
indicating that the object $obj$ possesses the attribute, 
we will add it to the set $A_{bt}$ in the operation \textit{op2.2}.
Otherwise, we shall add the attribute to the set $A_{nbt}$ as this object does not contain it.

\textbf{Example for $SC$ construction.}
    We pass in two Python objects \texttt{[1,2,3]} and \texttt{"abc"} for the native method \texttt{numpy.power}, 
    and we can retrieve call logs when executing the C function in Figure~\ref{motivating} as the follows.
    \begin{lstlisting}
API(o2, 'Chk', Long, False)
API(o2, 'Chk', Float, False)
API(o2.'Get', '__index__', False)
\end{lstlisting}

    We add the object \texttt{o2} to the $SC$s as $o1$ before calling this native method.
    When dealing with the calls on lines 1 and 2, we use operation \textit{op1.2} to update the $SC$:
    $ o2 \mapsto (T_{o2\_nbt} = \{float, long\}, A_{o2\_nbt} = \{\_\_index\_\_\}) $.

    For the call on line 3, we first add the attribute to $SC$s with the operation \textit{op2.3}.
    After line 3, $SC$s are: \\
    $ o2 \mapsto ( T_{o2\_nbt} = \{float, long\}, A_{o2\_nbt} = \{\_\_index\_\_\} ) $ \\
    $ \wedge o1.\_\_index\_\_\ \mapsto () $.
        
\subsection{Reverse}
    \label{sec: mutation}
    The purpose of reverse is to flip the result of the current branch condition 
    and then create a new Python object to fulfill the opposite path conditions.
    We select to build the original $SC$ and its reverses simultaneously.
    
    As indicated in Section~\ref{sec: construction}, 
    we traverse the call logs and use the operations in Table~\ref{table: constructing} to create the original $SC$.
    Before we update the original $SC$ for each operation, 
    we produce one reversed $SC$ by making a copy and adding the $val_{act}$ to the opposite set.
    Thus, a reversed $SC$ can represent the constraints from the native method entrance to the current branch condition, and another path to the current branch condition.

    The specifics of the reverse process are present in Algorithm~\ref{algo: mutation}.
    According to the API call, 
    the \texttt{get\_operates} function on line 4 obtains operations from Table~\ref{table: constructing}.
    From lines 5 through 16, we go over each operation.
    Line 6 generates a copy before updating the original $SC$, 
    while function \texttt{apply} in line 7 changes the original $SC$ depending on the operation.
    Lines 8-15 update the set against the operation in the copied $SC$.
    If this reversed $SC$ does not appear before, we include it in the result \texttt{mut\_list}. 
    \begin{algorithm}[h]
    \caption{Reverse on one $SC$}
    \label{algo: mutation}
    \KwIn{API function call logs \texttt{api\_list}}
    \KwOut{One original $SC$ \texttt{cons}, a set of reversed $SC$s \texttt{mut\_list}}
mut\_list $\gets$ list()

cons $\gets$ dict() 

\For(){api $\in$ api\_list}{
    ope\_list $\gets$ get\_operates(api)
    
    \For(){ope $\in$ ope\_list}{
        mut\_cons $\gets$ copy(cons)

        apply(cons, ope)

        \uIf(){ope is OP1.1}{
            update(mut\_cons, ope, $T_{nbt}$)
        }
        \ElseIf(){ope is OP1.2}{
            update(mut\_cons, ope, $T_{bt}$)
        }
        \ElseIf(){ope is OP2.2}{
            update(mut\_cons, ope, $A_{nbt}$)
        }
        \ElseIf(){ope is OP2.3}{
            update(mut\_cons, ope, $A_{bt}$)
        }
        \Else{
            apply(mut\_cons, op2.1)
        }

        mut\_list.add(mut\_cons)
    }
}
return cons, mut\_list
\end{algorithm}

    Because each rule only increases one constraint, 
    and we create one reversed $SC$ for each rule,
    the number of reversed $SC$ does not exceed the number of constraints in the original $SC$.

    \textbf{An example for the reverse.}
        Consider the API function calls in the example for $SC$ construction.
        When we process the call on line 1, 
        we add \texttt{long} to the $T\_{nbt}$ in the original $SC$, while building a reverse $SC$:
        $ o1 \mapsto ( T_{o1\_bt} = \{long\} ) $.

        After we process the calls on lines 2 and 3, we can produce two reversed $SC$s.
        The first is: 
        $ o1 \mapsto ( T_{o1\_bt} = \{float\}, T_{o1\_nbt} = {long} ) $,  \\
        and the second is: 
        $ o1.\_\_index\_\_ \mapsto () $  \\
        $ \wedge o1 \mapsto ( T_{o1\_nbt} = \{long, float\}, A_{o1\_bt} = \{\_\_index\_\_\} ) $.
        
\subsection{Generation}
    Algorithm~\ref{algo: generation} depicts the generation process.
From a high level, we first produce a piece of class definition code that causes the class to inherit from the types in the $T_{bt}$ but not in the $T_{nbt}$.
Then we can instantiate this class and modify the attributes based on $A_{bt}$ and $A_{nbt}$.
\begin{algorithm}[h]
    \caption{Generation $SC$}
    \label{algo: generation}
    \KwIn{One $SC$ \texttt{StrucCons}, the value set \texttt{ValueSet}}
    \KwOut{One Python object \texttt{obj}}
gen\_set $\gets$ topological\_sort(StrucCons).keys())

attrimap $\gets$ dict()

\For(){attri $\in$ gen\_set}{
    cons $\gets$ StrucCons[attri]

    builtins $\gets$ get\_builtin\_types()
    
    \uIf(){cons.$T_{bt} \neq \emptyset$}{
        inherits $\gets$ (builtins - cons.$T_{nbt}$) $\cap$ cons.$T_{bt}$
    }
    \Else(){
        inherits $\gets$ (builtins - cons.$T_{nbt}$)
    }

    self\_cls $\gets$ build\_class(inherits)

    up\_class(self\_cls, cons, attrimap, ValueSet)

    obj $\gets$ self\_cls()

    up\_object(obj, cons, attrimap, ValueSet)

    attrimap[attri] $\gets$ obj
}

return attrimap[attriset.last]

\end{algorithm}

Because $SC$ imposes constraints on both the object and its attributes, 
we must first generate attributes before producing the object. 
In line 1, we topologically order all keys in $SC$ based on their containment relationship.
For example, if object \texttt{o1} appears in $A_{bt}$ or $A_{nbt}$ of object \texttt{o2}, we consider \texttt{o2} contains \texttt{o1} and sort \texttt{o2} behind \texttt{o1}.
We use a map \texttt{attrimap} in line 2 to store all the attributes we have produced. 

The two language features \textit{dynamic attribute} and \textit{dynamic type} cannot be utilized to generate objects since the API functions only check for types and member methods based on the class definition code.
Thus, each object generation is completed in three phases.
First, we construct a class that inherits needed types from lines 5–10.
Then, in line 11, following the attribute constraints, 
we change the member variables and methods in the class definition code.
Finally, if the object is of collection types, as demonstrated in lines 12 to 13, 
we instantiate this class and adjust the elements depending on attribute constraints.

We compute the types that the class needs to inherit from lines 5 to 9,
where we subtract the types in the set $cons.T_{nbt}$ from the Python built-in type set \texttt{builtins}, 
then intersect the result with $cons.T_{bt}$.
If the set $cons.T_{bt}$ is empty, 
we do not conduct the intersection operation since the object is not required to inherit from a specific type.
In line 10, 
we produce a piece of class definition code to make the self-defined class inherits from the types in the set \texttt{inherits} in the function \texttt{build\_class}.

We change member methods and variables in the self-defined class 
depending on $cons.A_{bt}$ and $cons.A_{nbt}$ in the function \texttt{up\_class} on line 11.
We extract a member method or variable from the \texttt{attrimap} 
and add it to the class definition code if this attribute is required to exist.
For example, if the set $cons.A_{bt}$ is $\{\_\_index\_\_\}$,
we can add a method named \texttt{\_\_index\_\_} and a member variable 
as this method's return value to the class definition code.
The member variable's name can be generated randomly 
while its value comes from the value set \texttt{ValueSet}.

We instantiate this class on line 12.
The object \texttt{obj} is then updated at line 13 via the function \texttt{up\_object} 
if it is of collection types and there exist constraints on its elements,
where we add required attributes to this object.

In line 14, we append the \texttt{obj} object into the \texttt{attriset}.
The input object we produced for the $SC$ is the last element added to the \texttt{attriset} after finishing the loop from lines 4 to 14.

\textbf{An example for the generation.}
    The following $SC$ is used as an example to demonstrate the generating process: \\
    $ o1 \mapsto (T_{o1\_bt}=\{dict\}, A_{o1\_bt}=\{\_\_index\_\_, names\}) $  \\
    $ \wedge o1.\_\_index\_\_ \mapsto (T_{o1.\_\_index\_\_\_bt}=\{float\}) $  \\ 
    $ \wedge o1.names \mapsto (T_{o1.names\_bt}=\{list\}) $.  \\
    We start by sorting the three keys in the $SC$, and we can retrieve \texttt{gen\_set} as \texttt{o1.\_\_index\_\_}, \texttt{o1.names}, and \texttt{o1}, which implies we need to produce two attributes before building object \texttt{o1}. 
    \begin{lstlisting}[language=Python]
class self_class(dict):
  retvalue = 1.0
  def __index__(self):  return self.retvalue
obj1 = self_class()
obj1["names"] = []
\end{lstlisting}
    
    We produce a method \texttt{\_\_index\_\_} for the key \texttt{o1.\_\_index\_\_}.
    Because $T_{o1.\_\_index\_\_\_bt}$ contains $float$, this method must return an \texttt{float} value.
    Thus, we generate a member variable of type \texttt{float} and return it in this method.
    
    If the attribute \texttt{o.names} exists in the \texttt{attriset}, 
    we can utilize the matching value to produce a key-value pair and add it to the object \texttt{obj}.
    Or we can randomly generate a value to produce the key-value pair.
    We generate an empty list for constraints on \texttt{o1.names}, 
    as required by $T_{o1.names\_bt}$.
    
    For constraints on \texttt{o1}, 
    we first create a class that inherits from type \texttt{dict} based on its $T_{bt}$, 
    then add member method and variable to the class definition code (lines 2-3), 
    and last instantiate the class and insert the required key-value pair (lines 4-5). 
    After this, object \texttt{obj1} is the object we generate for this $SC$.
    \vspace{-0.3cm}
        
    \section{Evaluation}
        \subsection{Evaluation Setup}
    We instrument Python 3.8.5 and develop PyCing with Python 3.
    All of the code and data are open to the public
    ~\footnote{\url{https://anonymous.4open.science/r/pycing-present-6BC7}}.

    We traverse a list of the top 50 Python projects on GitHub by star rating.
    As the target of PyCing is testing Python native libraries, 
    We choose projects based on the three criteria listed below:
    (1) incorporate native libraries, 
    (2) do not use lazy evaluation methods~\cite{ZhangS21} since we cannot guarantee that a native method runs or only returns a symbol, and 
    (3) have bug reports about native libraries.
    We remain six projects from the fifty for our experiments.

    In contrast to statically typed languages that determine the method types and definitions during compilation, 
    obtaining native methods can be challenging as Python only determines the method definitions when executing.
    Besides, no type inference tools support method types~\cite{SaifullahAR20, Khan21, monat2020static}. 

    Thus, we opt to collect native methods from two sources: 
    existing studies and project documents.
    Several studies have analyzed native libraries for some Python projects~\cite{Tan0LRSS021, ZhangS21, HuZHX21}.
    We can glean native methods from their publicly available experimental data.
    Besides, we review the project documentation for all available methods and then verify their implementation to collect native methods.

\subsubsection{Reserch Questions}
    We propose three research questions, and by answering them, we will present the results of our experiment.
    \begin{itemize}[topsep=1pt, partopsep=1pt, leftmargin=15pt]
        \item \textbf{RQ1.} What is the efficiency of PyCing?
        \item \textbf{RQ2.} What is the test effectiveness of PyCing in real-world Python projects?
        \item \textbf{RQ3.} What is the test effectiveness of PyCing when compared to other fuzzers?
    \end{itemize}

    In answer to the first question, we evaluate the impact of PyCing instrumentation and settings on test efficiency.
    To answer the second question, we collect modern Python projects from GitHub
    and compare test cases in these projects with PyCing.
    Because existing tools cannot obtain code coverage from these projects, 
    we compare PyCing against two state-of-the-art Python fuzzers on a set of handwritten benchmarks to answer the third question. 

\subsubsection{Evaluation Criteria}
    As indicated in Section~\ref{sec: introduction}, native libraries do not provide code coverage.
    Thus, we evaluate testing effectiveness from 2 factors: 
    the number of $SC$s and the types of API functions.
    When the number of test cases is identical, 
    the number of $SC$s can indicate the number of examined execution paths, 
    while the types of API functions can reflect the structure coverage that native libraries can handle.

\subsubsection{Configuration}
    All experiments run on a PC equipped with an Intel Core i7 3.60GHz CPU and 32 GByte RAM. 
    The GitHub projects are built on Ubuntu 18 using Python 3.8.5. 
    We establish a time restriction of 24 hours for each project, 
    determined by the comparative experiments specified in papers or extra materials of existing Python testing tools.

\subsection{Performance of PyCing (RQ1)}
    Two aspects will influence the efficiency of PyCing:
instrumentation to the interpreter and configurations in the generation phase of PyCing.

We first assess the influence of instrumentation on execution performance.
The instrumented interpreter is compared to a standard interpreter and a commonly used dynamic binary instrumentation framework \textit{pin}~\cite{LukCMPKLWRH05}.
We utilize the test cases in these projects as benchmarks since they can represent common usage, 
and we measure the execution time with the testing framework \textit{Pytest}.
Because \textit{Pytest} supports multi-processing,
we can run test cases with different numbers of processes.
We can directly invoke \textit{Pytest} as a Python module for the two interpreters.
For the framework \textit{pin}, we use the default \textit{pintool} to monitor at the function level and execute on the standard interpreter inside the framework.
Project \textit{CPython} is excluded from this experiment as many native methods in this project deal with suspended execution or multi-processing, both of which have a substantial impact on the running time count.

Table~\ref{table: runtime} shows the number of test cases and execution times under a single process (1), 
two processes (2), and four processes (4). 
If one test case runs for more than an hour or the operating system terminates the Python virtual machine, 
we will mark it as \textbf{NA} for this configuration.
According to the data, our instrumentation strategy has an acceptable influence on execution performance,
while dynamic binary instrumentation increases the time by 5x to 10x.
Increasing the number of processes might lower the time under dynamic binary instrumentation.
However, testing one native method requires several rounds of sequential execution, 
and increasing the number of processes has little effect on test efficiency.
In the \textit{SciPy} project, the \textit{pin} fails 
because of conflicts when instrumenting one binary file under the multiprocess mode, 
or it takes too much memory for storing interim results, causing the OS to terminate it.
\begin{table}[h]
    \small
    \centering
    \setlength\tabcolsep{1.7pt}
    \renewcommand{\arraystretch}{1.5}
    \caption{Execution time under different instrumentation strategies (seconds).}
    \begin{tabular}{ccccc|ccc|ccc}
        \toprule
        \multirow{2}*{Project} &  \multirow{2}*{Tests}  & 
                                  \multicolumn{3}{c|}{Standard} & 
                                  \multicolumn{3}{c|}{Instrumented} & 
                                  \multicolumn{3}{c}{Pin} \\
        \cline{3-11}
                && 1 & 2 & 4 & 1 & 2 & 4 & 1 & 2 & 4   \\
        \hline
        Matplotlib  &  8K    &  404  &  243  &  160  &  434  &  297  &  183  &  2,762  &  1,489  &  991  \\
        NumPy       &  17K   &  151  &  107  &  73   &  179  &  121  &  89   &  1,034  &  667   &  446  \\
        Pandas      &  132K  &  1,644 &  969  &  787  &  1,992 &  1,171 &  893  &  9,653  &  5,476  &  NA   \\
        Pillow      &  2K    &  49   &  25   &  16   &  52   &  29   &  18   &  350   &  217   &  157  \\
        SciPy       &  35K   &  477  &  305  &  245  &  545  &  333  &  255  &  NA    &  NA    &  NA \\  
        \bottomrule
    \end{tabular}
    \label{table: runtime}
\end{table}

To prevent combinatorial explosions, we restrict the number of $SC$s in each loop's generation phase.
We choose a value every 200 as the upper limit in the range of 200 to 1200 to find an appropriate configuration.
PyCing can complete in under 2 hours in each setting.
Due to space limitations, 
we only display the data from the project \textit{NumPy} and place the results from other projects on our website.

\begin{figure}[h]
    \centering  \includegraphics[width=\linewidth]{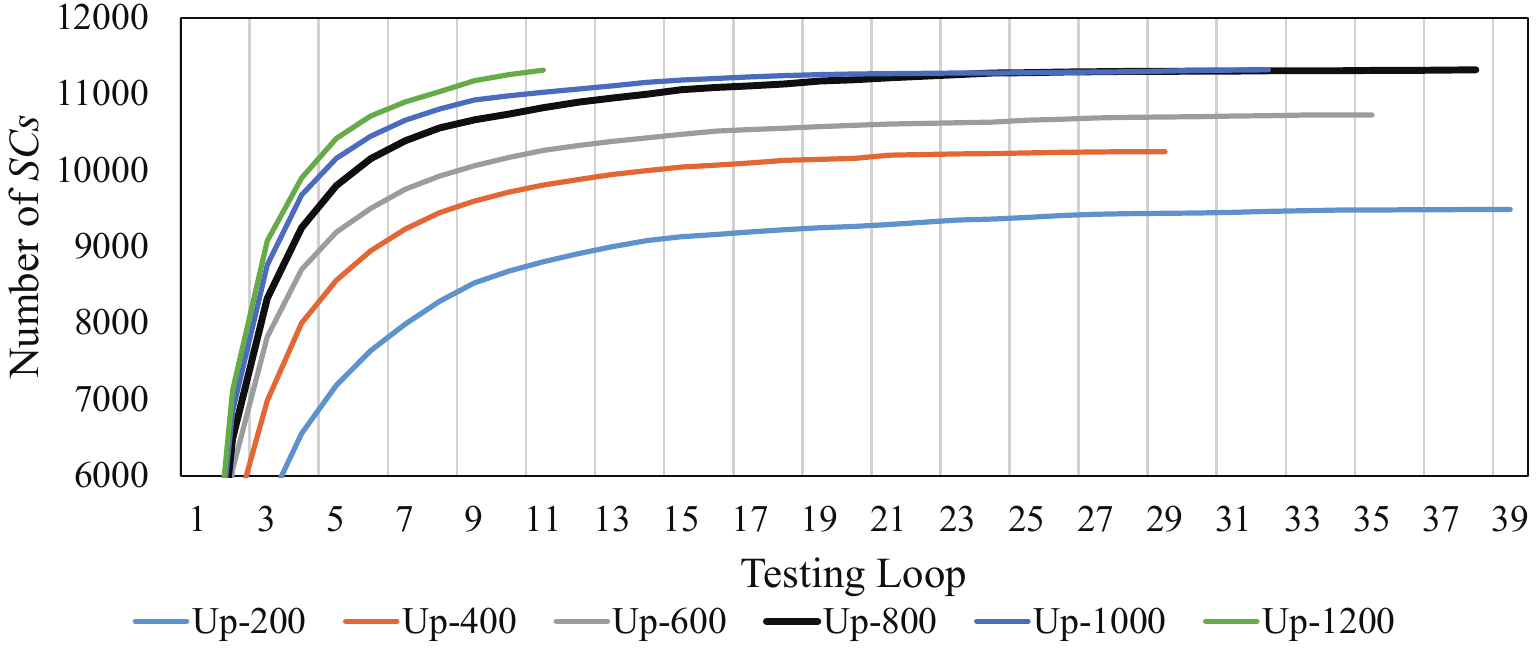}
    \caption{Number of explored $SC$s under each upper limit.}  \label{limit compare}
\end{figure}
The results in Figure~\ref{limit compare} indicate that increasing the upper limit of $SC$s helps us to reach the top bound faster, and the exploration can finish in less than 40 loops.
However, raising the maximum limit needs more time to complete one loop.
Even though PyCing can complete exploration in 12 loops under the 1200 upper limit, this configuration takes longer than others.
Considering the resource usage and execution speed, 
we believe 800 $SC$s per loop is the acceptable configuration.

\textbf{Conclusion.}
The execution time comparison reveals that our instrumentation strategy has less influence on execution efficiency.
Obtaining execution information from native libraries via dynamic binary instrumentation can be time-consuming or may fail since it demands monitoring the whole Python virtual machine.

\subsection{Test Effectiveness of PyCing (RQ2)}
    In answering this question, 
we will first evaluate the efficiency of test cases in real-world projects, 
then investigate the improvement made by PyCing.

To quantify the size of native libraries developed by these projects, 
we use the tool \textit{cloc} to count the lines of source code in C/C++, FORTRAN, and Cython.
Besides, we count the types and numbers of Python/C APIs employed in the native libraries.
Columns 3-6 of Table~\ref{table: line of code} demonstrate that the six projects develop their native libraries using a variety of languages and a large number of API functions, 
some of which are multilingual, providing challenges for building summaries.
\begin{table}[h]
    \centering
    \small
    \setlength\tabcolsep{4pt}
    \renewcommand{\arraystretch}{1.5}
    \caption{Basic information of experimented projects.}
    \begin{tabular}{cccccc}
        \toprule
        Project  &  Version  &  C/C++  &  FORTRAN  &  Cython  &  API  \\
        \hline
        CPython          &  3.10.1   &  552K   &  0     &  0       &  6,249   \\
        Matplotlib       &  3.5.1    &  69K    &  0     &  0       &  56      \\ 
        NumPy            &  1.22.3   &  185K   &  416   &  3,589   &  5,885   \\
        Pandas           &  1.4.2    &  7,017  &  0     &  20K     &  38,943  \\
        Pillow           &  9.0.1    &  33K    &  0     &  0       &  136      \\
        SciPy            &  1.8.0    &  135K   &  74K   &  11K     &  31,131  \\ 
        \bottomrule
    \end{tabular}
    \label{table: line of code}
\end{table}

We first count the number of $SC$s explored by existing test cases.
The number of native methods is shown in the second column of Table~\ref{table: origin test case}, 
and the number of types of input data used by test cases to the native methods is listed in the third column.
The fourth column shows the number of $SC$ explored by the test cases.
According to the results, the test cases employ a wide range of input data to test native methods.
But when compared with the number of explored $SC$s, 
we can observe that these test cases fall short of addressing execution paths with structure constraints.
\begin{table}[h]
    \centering
    \small
    \renewcommand{\arraystretch}{1.5}
    \caption{Number of $SC$s covered by existing test cases.}
    \begin{tabular}{cccc}
        \toprule
        Project     &  Native Method  &  Input data  & $SC$  \\
        \hline
        Cpython     &  499   &  176K    &  1,043  \\
        Matplotlib  &  10    &  1,889   &  10     \\
        NumPy       &  194   &  185K    &  218    \\
        Pandas      &  188   &  467K    &  334    \\
        Pillow      &  16    &  1,475   &  35     \\ 
        SciPy       &  224   &  33K     &  242    \\
        \bottomrule
    \end{tabular}
    \label{table: origin test case}
\end{table}

\begin{figure}[h]
    \centering  \includegraphics[width=\linewidth]{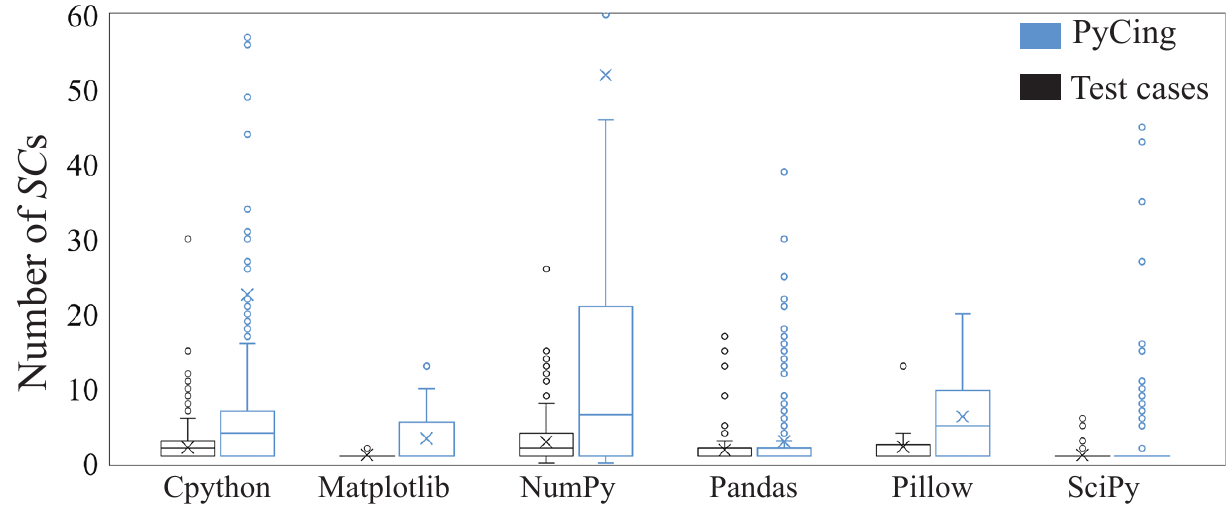}
    \caption{Number of $SC$s explored by test cases and by PyCing.}  \label{project compare}
\end{figure}

We compare the number of $SC$ explored by the test cases to the number identified by PyCing to explore whether PyCing can increase test effectiveness.
Figure~\ref{project compare} shows the comparing results.
Because PyCing finds a significant number of $SC$s for a few native methods, 
we exhibit the distribution based on the number of $SC$s explored by the test cases.

As the results show, 
PyCing can find more $SC$s than test cases for most native methods, 
particularly in the \textit{SciPy} project, 
where test cases examined 242 $SC$s whereas PyCing examined 160K totally.

As mentioned in Section~\ref{sec: instrument}, 
we instrument type checking and attribute extraction API functions. 
To explore which category of API functions aid PyCing to improve test effectiveness, 
we count two kinds of API functions covered by PyCing and test cases.
Besides, because the return values of API functions can cause the code to run alternative branches, 
we count the different types of return values of these functions.
In particular, we regard returning true and false as two types of return values for type checking API methods and returning NULL and non-NULL to be two types for attribute extraction API functions.

\begin{figure}[h]
    \centering  \includegraphics[width=\linewidth]{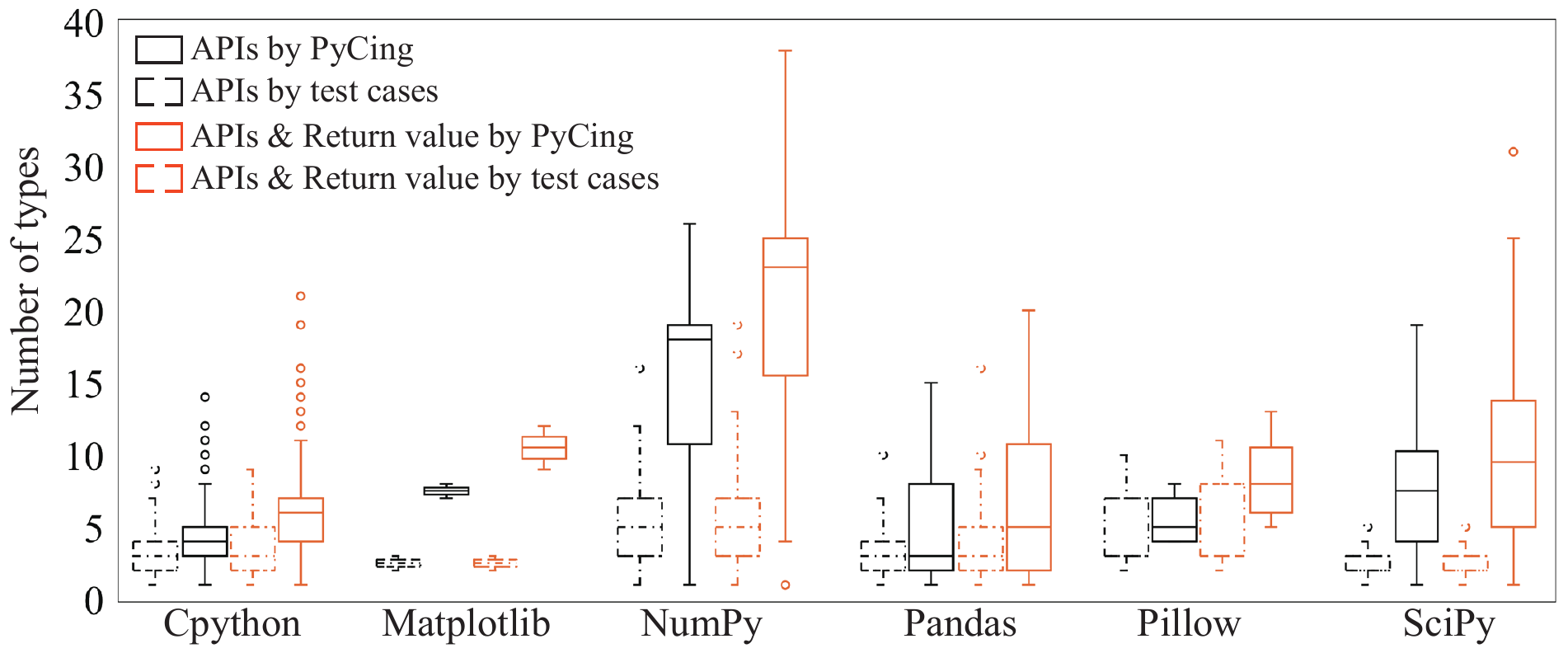}
    \caption{Number of types of API functions and return value covered by test cases and PyCing.}  \label{table: type cond}
\end{figure}

Figure~\ref{table: type cond} presents the results.
As shown, PyCing can cover more APIs for type checking and attribute extraction with the same amount of test cases, 
particularly for attributes.
Moreover, PyCing can result in more types of return values for these API calls.

We also consider the influence of structures not covered by the test cases on the test effectiveness.
The second column in Table~\ref{table: newly structure} indicates the new API functions discovered by PyCing, and the third column gives the number of native methods that utilize these API functions.
In the table, API functions ending in \texttt{Exact} return true if the argument is of the checked type but not a subtype, while API functions ending in \texttt{get} extract attributes.
The results in Figure~\ref{table: type cond} and Table~\ref{table: newly structure} demonstrate that 
PyCing can improve test effectiveness in two ways.
First, PyCing can trigger various API return values for the API functions examined by the test cases.
Second, PyCing can cover API functions not covered by test cases.
\begin{table}[h]
    \centering
    \small
    \renewcommand{\arraystretch}{1.5}
    \caption{New API functions explored by PyCing.}
    \begin{tabular}{m{1.1cm}m{5cm}c}
        \toprule
        Project  &  Structures  & \# Met  \\
        \hline

        CPython  &  set, object\_getitemstring  &  2  \\

        \hline
        
        Matplotlib  &  bytes, dict, list, sequence, tuple, tupleExact, sequence\_getitem,  &  2  \\

        \hline
        
        NumPy  &  bytearray, complexExact, dictExact, list, mapping, mapping\_getitemstring, object\_getitem  &  94  \\

        \hline

        Pandas  &  object\_getattr, object\_getitemstring  &  5  \\

        \hline

        Pillow  &  object\_getattr  &  3  \\

        \hline

        SciPy  &  bytearray, complex, floatExact, tupleExact, object\_getattr  &  12  \\

        \bottomrule
    \end{tabular}
    \label{table: newly structure}
\end{table}

\subsubsection{Bug Detection Results}
    In this paper, we use PyCing to detect memory leaks and segmentation faults in native libraries.
The testing framework \textit{Pytest} provides a checker \textit{pytest-leaks}~\cite{pytestleaks} for detecting memory leaks,
while segmentation faults will make the interpreter terminated.
We first use PyCing to produce test cases for the native methods in the six projects,
and run these test cases via \textit{Pytest}.
We can produce test cases to trigger memory leaks in 9 native methods, and segmentation faults in 10 methods.
Table~\ref{leaks} shows distributions of bugs among the projects.
\begin{table}[h]
    \centering
    \small
    \renewcommand\arraystretch{1.5}
    \caption{Bugs detected by PyCing} 
    \begin{tabular}{ccc}
        \toprule
        Projects &  Memory Leaks  &  Segmentation Faults  \\
        \hline
        Cpython      &  1  &  1  \\
        Matplotlib   &  1  &  -  \\
        NumPy        &  4  &  2  \\
        Pandas       &  -  &  2  \\
        Pillow       &  -  &  3  \\
        SciPy        &  3  &  2  \\
        \hline
        \textbf{SUM} &  \textbf{9}  &  \textbf{10}  \\
        \bottomrule
    \end{tabular}
    \label{leaks}
\end{table}

Based on leaks we report to the developers, 
we can find that developers often cause leaks when using API functions to extract values from input data while ignoring the side effects of the change in the reference count of the return value.
The following code utilizes two APIs to obtain the first element from an object \texttt{o} of type \texttt{list}.
However, the first API raises \texttt{item1}'s reference count.
Thus, when developers forget to reduce the reference count, a leak occurs. 
We advise developers to use uniform API categories to minimize misunderstandings about the behavior of the returned response.
\begin{lstlisting}[language=C]
PyObject item1 = PySequence_GetItem(o, 0);  // increase
PyObject item2 = PyList_GetItem(o, 0);  // not increase
\end{lstlisting}

We cannot pinpoint the source of segmentation issues.
It is due to the lack of debugging information in the project compilation process.
Besides, the OS stops the interpreter before it outputs error information.
As a result, even if we may produce many test cases that trigger the bugs, we count these bugs at the native method level.

\textbf{A case study.}
The following code shows a leak found by PyCing in the project \textit{CPython} and it affects Python 3.9, 3.10, and 3.11~\cite{aleak}.
When the native libraries \texttt{\_flatten1} from lines 8 to 12 returns 0, function \texttt{\_tkinter\_flatten} does not reduce the reference of \texttt{obj} before returning at line 5.
The input object needs to execute the false branch at line 9 to trigger this leak, and PyCing produces a \texttt{str} object for it.
As test suites do not contain such a type, input data from existing fuzzers can not cover the satisfying type to trigger it.
The input object generated by PyCing has been added to the test suites of this project.
\begin{lstlisting}[language=C]
PyObject * _tkinter_flatten(PyObject * obj){
    ...
    if(!_flatten1(obj)){
        // fix patch: + Py_XDECREF(obj)
        return 0;      }
}
int _flatten1(PyObject * obj){
    if(PyTuple_Check(obj) || PyList())
        return 1;    
    return 0;
}
\end{lstlisting}

\textbf{Conclusion.}
Real-world Python projects include plenty of native libraries written in multiple statically typed programming languages.
These native libraries use a plethora of Python/C APIs in condition branches to handle various kinds and attributes of input objects.
When comparing the number of APIs used to the number of APIs covered by original test cases, we can find that native libraries are not well tested, even though these projects have a considerable number of test cases and code coverage of 95\% in Python.

The results of six real-world experiments reveal that PyCing can effectively explore execution paths, 
especially those related to the types and attributes of input objects.
PyCing can aid in the automated generation of test cases for native library testing and memory leak detection.

\subsection{The Comparison with other Fuzzers (RQ3)}
    To evaluate the effectiveness of PyCing, we choose to compare it with two state-of-art Python fuzzers, 
\textit{pythonfuzz}~\cite{pythonfuzz} and \textit{python-afl}~\cite{pythonafl}.
\textit{pythonfuzz} is a coverage-guided fuzzing tool maintained by Google, 
which detects unhandled exceptions and memory limitations in Python libraries, 
while \textit{python-afl} applies \textit{American Fuzzy Lop (AFL)} for pure Python code. 
Currently, \textit{python-afl} is an experimental module.  

As indicated in Section~\ref{sec: introduction}, 
current test tools cannot acquire code coverage from native libraries in the experimental projects, 
and the two Python fuzzers are unable to exploit structure constraints.
To compare two fuzzers with PyCing fairly, 
we generate multiple sets of codes that contain the type and attribute judgments as benchmarks.
Because \textit{pythonfuzz} and \textit{python-afl} are for pure Python and PyCing for testing native libraries,
each set contains C and Python code snippets with the same functionality.

To ensure that Python and C code behave identically, we use Python/C APIs to process the types and attributes of input objects in C code, then write code templates with the same functionality for each API function in Python code.
A type-checking API function is equivalent to the built-in method \texttt{isinstance}, 
and an attribute-extracting API function corresponds to another built-in method \texttt{in}.
For simplicity, we do not process the attributes of the input objects in the benchmarks.

Thus, we can generate C code by adding multiple branches and randomly selecting API functions in the branch conditions, 
where we ensure each function receives the same object.
Because the attribute-extracting API functions do not return a boolean variable, 
we compare the return result to \texttt{NULL} in the branch condition. 
Then, in the Python code, we may retain the form of the \texttt{if} statements while substituting handwritten templates for the APIs.

\begin{figure}[h]
    \centering
    \includegraphics[width=\linewidth]{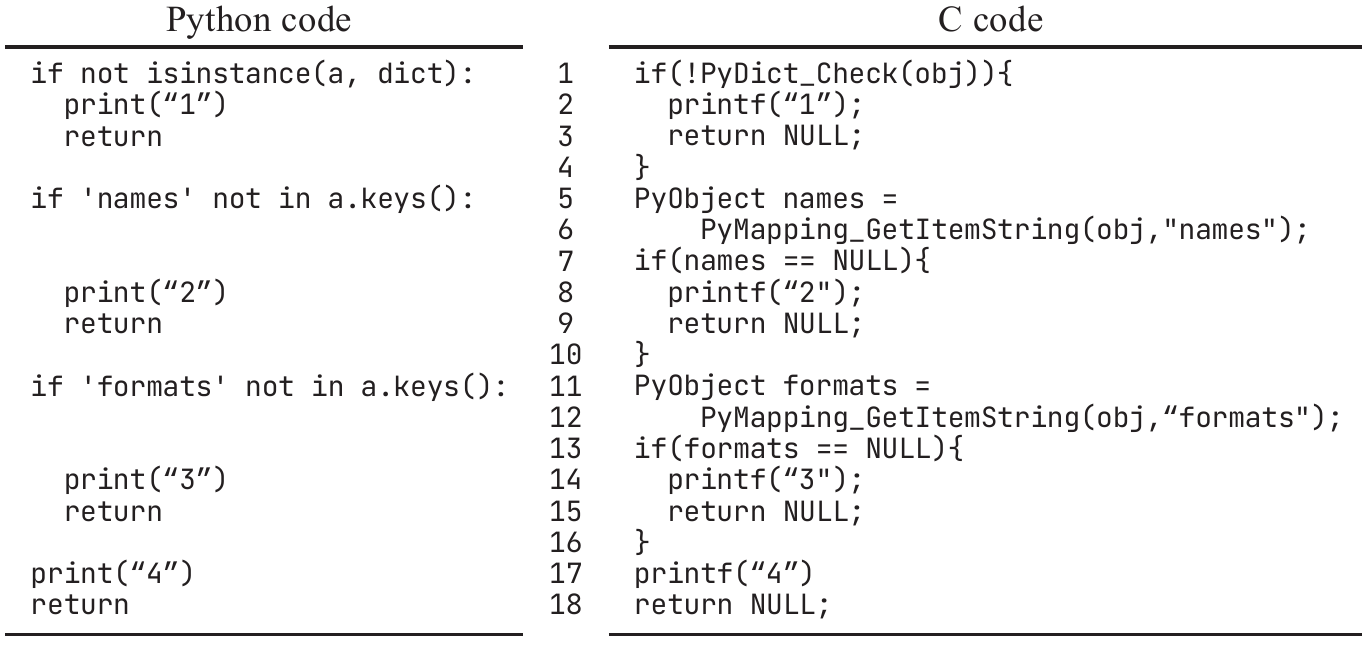}
    \caption{Two snippets of handcrafted code for testing \textit{Pythonfuzz} and \textit{Python-afl}.} 
    \label{handcrafted code}
\end{figure}

Figure~\ref{handcrafted code} presents two pieces of codes.
The left piece of code is for \textit{python-afl} and \textit{pythonfuzz}, while the right is for PyCing.
In both Python and C code, lines 1 to 4 check if the input is of the \texttt{dict} type. 
Lines 5 to 10 in both sides inspect if the object includes the key \texttt{'names'}, and lines 11 to 16 check for the key \texttt{'formats'}.
We use the Python code as fuzz target for \textit{Pythonfuzz} and \textit{python-afl} while compiling the C code to produce dynamic link libraries for PyCing.

\textit{pythonfuzz} and \textit{python-afl} can only enter the True branch of the first boolean statement.
The two tools are not effective in dealing with constraints on the structures.
PyCing can explore all the execution paths based on the output of the handcrafted code.
When the input objects generated by PyCing are seeded to these tools, 
the tools can mutate the values in the objects, covering the paths in these benchmarks.

\textbf{Conclusion.}
Existing fuzzers are more concerned with the value of the input data than with its structure; as a result, they are ineffective when an execution path incorporates input data structures.
Adding support for complicated structures and using feedback other than code coverage to assist generation can be improvable.

    \section{Threats to Validity}
        We will discuss some internal threats and external threats to our work in this section.

\textbf{Internal Threats.} 
The native method collection is one of the most significant internal threats.
There are some native methods not found by the relevant research, or not recorded in the project documentation.
But we believe that there are fewer undocumented native methods, or that some are soon to become outdated.
On the other hand, the project does not want these methods to be called directly.

Another threat arises with the generation of new input data.
Because we do not employ a constraint solver in the generation, there exist some constraints that we can not create new objects.
Besides, because PyCing primarily focuses on the types and attributes, 
it cannot handle numerical constraints between two objects.
However, embedding the structure constraints into a solver can result in deviation, 
and a solver may slow down the fuzzing process.
Value constraints between attributes are difficult to find and handle even by analyzing the source code.

\textbf{External Threats.} 
We exclusively use PyCing to identify memory leaks, and PyCing may not perform well on other defects. 
Besides, because the memory leaks discovered by us came from four Python projects, the results may not apply to other projects.
Also, we only build PyCing and instrument APIs with \textit{CPython}, and may not work on other implementations of Python or native libraries.
But, the six projects we choose are essential components in virtually all Python code, 
and the experiment results show that PyCing can investigate more execution paths in native libraries.
The proposed instrumentation method can also be applied to other interpreters.

    \section{Related Works}
        In this section, we briefly describe some related works.

\textbf{Detecting Memory Leaks via Testing Approaches.} 
Some studies employ testing techniques such as fuzzing 
to uncover new paths and provide methodologies for monitoring memory usage and detecting memory leaks.
\textit{Valgrind}~\cite{NethercoteS07} is a widely used tool for detecting memory leaks. It monitors the execution of programs to collect information on memory allocation and free.
Cristian et al.~\cite{CadarDE08} use a solver on the acquired path conditions to generate new input data.
Ma et al.~\cite{JunSYTJ17} employ UI testing techniques and dump heap files to evaluate memory leaks. 
For 10 open source apps and 35 commercial apps, they report more than one leaked action or fragment.
They create test cases by defining test generating methods for two types of GUI event sequences and following certain typical leak patterns.
Wu et al.~\cite{WuLXGZYZ16} combines static analysis and model checking to detect resource leaks in Android Apps.
However, Their test scenarios do not involve cross-language situations.

\textbf{Cross-language Analysis.} 
Some researchers attempt to provide summaries for native libraries to guide native method testing.
Fourtounis et al.~\cite{FourtounisTS20} expand the Java call graph to include both Java and native code. 
Zhao et al.~\cite{ZhaoWLPKW19} focus on the calls to several high-risk system APIs in native libraries and collect branch prediction data for the execution path that includes these APIs. 
To assist fuzzing in Java code, they employ path conditions gathered in native libraries.
Lee~\cite{Lee19} creates summaries for native libraries, which are subsequently used in the analysis of Java code.

\textbf{Testing Approaches in Python.}
Many studies use testing techniques to identify flaws or wasteful code in Python projects.
Lukasczyk et al.~\cite{LukasczykKF20} proposed a method for creating test cases automatically.
Holmes et al.~\cite{Holmes0BGZG20} put forward a method for guiding generation that relies on relative lines.
The two methods focus on value generation and require users to define the types.

    \section{Conclusion}
        In this paper, we propose a lightweight approach for testing Python native libraries, 
and we implement it in a tool called PyCing.
By imposing constraints on the structures of input data and instrumenting the Python Interpreter, 
PyCing can overcome the language barrier and improve test efficiency.
The experimental results in six real-world Python projects show that PyCing explores execution paths more efficiently than state-of-the-art Python fuzzers and can generate seeds for these fuzzers to boost their test effectiveness.
PyCing also finds memory leaks in 9 native methods and segmentation faults in 10 methods.
Developers have confirmed eight leaks.
In the future, we will introduce value constraints in native libraries to PyCing to improve test effectiveness.
    
    \clearpage
    \bibliographystyle{unsrt}
    \bibliography{references}

\end{document}